\documentstyle[12pt]{article}
\begin{document}
\def\thefootnote{\fnsymbol{footnote}}
\begin{flushright}
KANAZAWA-97-07  \\ 
April, 1997
\end{flushright}
\vspace*{2cm}
\begin{center}
{\LARGE\bf Effect on the electron EDM due to abelian gauginos 
in SUSY extra $U(1)$ models}\\
\vspace{1 cm}
{\Large Daijiro Suematsu}
\footnote[1]{e-mail:suematsu@hep.s.kanazawa-u.ac.jp}
\vspace {1cm}\\
{\it Department of Physics, Kanazawa University,\\
        Kanazawa 920-11, Japan}\\    
\end{center}
\vspace{1cm}
{\Large\bf Abstract}\\  
The electric dipole moment of an electron (EDME) is investigated in
the supersymmetric extra $U(1)$ models.
Neutralino sector is generally extended in these models and then
the neutralino contribution will be important for the analysis of the EDME.
Kinetic term mixings of abelian gauginos are taken into account in
our analysis.
Numerical results for the extra $U(1)$ models show that the EDME 
can be affected by the extra $U(1)$ in a certain range of soft 
supersymmetry breaking 
parameters even if the extra $U(1)$ gauge boson is heavy.
The EDME may be a clue to find an extended gauge structure in the
supersymmetric models.
\newpage
\setcounter{footnote}{0}
\def\thefootnote{\arabic{footnote}}
Recently the standard model(SM) has been confirmed in the incredible accuracy
through the precise measurement at LEP.
Nevertheless, it has still not been considered as the fundamental theory of
particle physics and physics beyond the SM is eagerly explored.
Along this line supersymmetrization of the SM is now considered
as the most promising extension\cite{n}.
Even in this minimal supersymmetric standard model(MSSM), however, 
there remain some theoretically unsatisfactory features in addition to
the existence of too many parameters.
One of these is known as the $\mu$-problem\cite{mu}. 
The MSSM has a supersymmetric
Higgs mixing term $\mu H_1H_2$. To cause an appropriate radiative
symmetry breaking at the weak scale we should put $\mu\sim
O(G_F^{-1/2})$ by hand, where $G_F$ is a Fermi constant.
Although in the supersymmetric model its typical scale is generally 
characterized by the supersymmetry breaking scale $M_S$, 
there is no reason why
$\mu$ should be such a scale because it is usually considered to be 
irrelevant to the supersymmetry breaking. 
The reasonable way to answer this issue is to consider the origin of 
$\mu$ scale as a result of the supersymmetry breaking\cite{musol}. 
One of such solution is the introduction of a
singlet field $S$ and replace $\mu H_1H_2$ by a Yukawa type coupling
$\lambda SH_1H_2$ \cite{singlet}. 
If $S$ get a vacuum expectation value(VEV) of order
1~TeV as a result of  radiative corrections to the soft supersymmetry 
breaking parameters\cite{rad}, 
$\mu\sim O(G_F^{-1/2})$ will be realized dynamically 
as $\mu=\lambda\langle S\rangle$.

The extra $U(1)$ models can be a simple candidate of this 
scenario\cite{sy,extra}.
Moreover, extra $U(1)$s often come from the high energy fundamental theory
like superstring\cite{string}.
It is very interesting subject to find a signature of such an 
extra gauge symmetry.
Recent LEP precise measurement and Tevatron direct search tell us that 
extra gauge bosons may be too heavy to find it in near future\cite{exp}.
In supersymmetric models there are gauginos which may be rather light
for certain range of soft breaking parameters
even if corresponding gauge bosons are heavy. In such a case
we may be able to find a clue of the additional gauge structure through the
superpartner sector. 
In the above mentioned extra $U(1)$ models 
a relevant noticable sector is the neutralino sector, which is  
extended  in comparison with that of the MSSM\cite{n} 
at least by two components, 
that is, an abelian gaugino and
a superpartner fermion of the singlet Higgs $S$.

There may be some interesting phenomena in which the neutralino sector
can play an important role.
Such a representative example is the electric dipole moment of an 
electron (EDME) and a neutron. 
In supersymmetric models it has been known that there are one-loop 
contributions to the EDME due to the co-operation of the superpartners 
( {\it e.g.} charginos, neutralinos and so on )
and the additional CP violating phases in the soft supersymmetry
breaking parameters.
In the MSSM a lot of works for the EDM of a neutron and an electron 
have been done\cite{edm,edm2}. 
Through such studies it has been suggested that the contributions from
charginos and neutralinos can become important
in the suitable parameter region.
In the case of EDME there are no contribution coming from 
colored fields\footnote{Here we assume that there is no R-parity
violating term.} 
and then the effect of the neutralino sector is
expected to appear clearly in its estimation. 
In the models we are considering here, 
there are some changes from the MSSM in the neutralino sector and the 
effects induced by this change may be able to appear in the EDME explicitly.
As the present experimental bound of the EDME\cite{expedm} is near the 
predicted value from the MSSM, 
it is very interesting to study the influence on the EDME which
comes from the existence of an extra $U(1)$ gauge boson.

In this letter we consider the minimally extended MSSM with an extra $U(1)_X$
and a gauge singlet Higgs $S$ which has a coupling $\lambda SH_1H_2$ 
with doublet Higgses $H_1$ and  $ H_2$. 
We note here an additional feature required in the treatment of 
the neutralino sector of the multi $U(1)$ models.
That is, there can be gauge invariant kinetic term mixings among abelian 
vector multiplets.\footnote{
It has been pointed out that
there can occur abelian gauge kinetic term mixings in the suitable 
models\cite{mixing,mixing1}.
Their supersymmetrization is represented by the mixings among
abelian vector multiplets. Some works on this phenomenological effects 
have been done by now\cite{mixing1,mixing2}.}
We derive the general formulus for the EDME in such models taking 
account of this gaugino mixing between $U(1)_Y$ and $U(1)_X$.
This formulus is investigated numerically 
and the effect of extra $U(1)_X$ symmetry on the EDME is discussed
to find some clue of the fundamental theory in high energy region.

In the general supersymmetric models with two abelian factor groups,
the supersymmetric gauge invariant kinetic terms of the abelian part 
for these are expressed by 
using chiral 
superfields $\hat W_\alpha^a$ and $\hat W^b_\alpha$ 
for $U(1)_a\times U(1)_b$ as
\begin{equation}
{1\over 32}\left(\hat W^{a\alpha} \hat W^a_\alpha\right)_F
+{1\over 32}\left(\hat W^{b\alpha} \hat W^b_\alpha\right)_F
+{\sin\chi\over 16}\left(\hat W^{a\alpha} \hat W_{\alpha}^b\right)_F,
\end{equation}
where $\hat W^a$ contains a gaugino $\lambda^a$, a gauge field
strength $F^a_{\mu\nu}$ and an auxiliary field $D^a$ as the component
fields.
These can be canonically diagonalized by using the transformation,
\begin{equation}
\left(\begin{array}{c} \hat W^a \\ \hat W^b \\ \end{array}\right)
=\left(\begin{array}{cc}1 & -\tan\chi \\ 0 & 1/\cos\chi \\ 
\end{array}\right)\left(\begin{array}{c} W^a \\ W^b \\
\end{array}\right).
\end{equation}
This transformation affects not only the gauge
field sector but also the sector of gauginos $\lambda_{a,b}$ and 
auxiliary fields $D_{a,b}$. 
The modification in the gaugino interactions can be summarized as
\begin{equation}
g_a^0Q_a\hat \lambda^a+g_b^0Q_b\hat
\lambda^b=g_aQ_a\lambda^a+\left(g_{ab}Q_a+g_bQ_b\right)\lambda^b,
\end{equation}
where $\lambda_{a,b}$ are canonically normalized gauginos.
$Q_a$ and $Q_b$ represent the charges of $U(1)_a$ and $U(1)_b$.
The couplings $g_a, g_{ab}$ and $g_b$ are related to the original ones
$g_a^0$ and $g_b^0$ as,
\begin{equation}
g_a=g_a^0, \qquad g_{ab}=g_a^0\tan\chi, \qquad g_b={g_b^0\over\cos\chi}. 
\end{equation}

Using this canonically normalized basis, we can write down the modified
quantities relevant to the neutralino sector in the present models.
Here we give the concrete forms of the
neutralino mass matrix and gaugino-fermion-sfermion interactions.
If we take the canonically normalized neutralino basis as 
${\cal N}^T=(-i\lambda_{W_3}, -i\lambda_Y, 
-i\lambda_X, \tilde H_1, \tilde H_2, \tilde S)$ and define their mass terms as
${\cal L}_{\rm mass}^{\rm n}=-{1\over 2}{\cal N}^T{\cal MN}+{\rm h.c.}$,
the 6 $\times$ 6 neutralino mass matrix ${\cal M}$ can be expressed as
\begin{equation}
\left( \begin{array}{cccccc}
M_W & 0 & 0 &m_Zc_W\cos\beta & -m_Zc_W\sin\beta &0 \\
0 & M_Y & C_1 & -m_Zs_W\cos\beta & m_Zs_W\sin\beta &0 \\
0 & C_1 & C_2 & C_3 & C_4 & C_5 \\
m_Zc_W\cos\beta &-m_Zs_W\cos\beta &  C_3& 0 & 
\lambda u & \lambda v\sin\beta \\
-m_Zc_W\sin\beta & m_Zs_W\sin\beta &C_4 & \lambda u &
0 & \lambda v\cos\beta \\
0 & 0& C_5 & \lambda v\sin\beta & \lambda v\cos\beta & 0\\
\end{array} \right),
\end{equation}
where $v$ and $u$ are defined as 
$v=(\vert\langle H_1^0\rangle\vert^2+\vert\langle H_2^0\rangle\vert^2
)^{1/2}$ and 
$u=\vert\langle S\rangle\vert$.
Matrix elements $C_1\sim C_5$ are affected by the 
kinetic term mixing. They are represented as\footnote{
In this expression we introduce the effect originated from the 
abelian gaugino mass mixing as $M_{YX}$, which may exist at the 
Planck and may also be yielded through the low energy quantum effects.}
\begin{eqnarray}
&&C_1=-M_Y\tan\chi +{M_{YX}\over\cos\chi},\quad
C_2=M_Y\tan^2\chi+{M_X \over
\cos^2\chi}-{2M_{YX}\sin\chi\over\cos^2\chi},  \nonumber \\
&&C_3={1\over \sqrt 2}\left(g_Y\tan\chi+{g_XQ_1\over\cos\chi}\right)
v\cos\beta,\quad
C_4={1\over \sqrt 2}\left(-g_Y\tan\chi+{g_XQ_2\over\cos\chi}\right)
v\sin\beta,\nonumber\\
&&C_5={1\over \sqrt 2}{g_XQ_S\over\cos\chi}u,
\end{eqnarray}
where $M_W, M_Y$ and $M_X$ are soft supersymmetry breaking gaugino masses.
$Q_1, Q_2$ and $Q_S$ are the extra $U(1)$ charges of Higgs
chiral superfields $H_1, H_2$ and $S$.
Neutralino mass eigenstates $\tilde\chi_i^0(i=1\sim 6)$ are related 
to ${\cal N}_j$
through the mixing matrix $U$ as
\begin{equation}
\tilde\chi_i^0=\sum_{j=1}^6 U_{ij}^T{\cal N}_j.
\end{equation}
The change in the gaugino interactions can be confined into 
the extra $U(1)_X$ gaugino sector and by using eq.(3) new interaction 
terms can be expressed as,
\begin{eqnarray}
&&{i \over \sqrt 2}\left[\tilde \psi^*\left(-g_YY\tan\chi+{g_XQ_X\over
\cos\chi}\right)\lambda_X\psi-\left(-g_YY\tan\chi+{g_XQ_X\over\cos\chi}
\right)\bar\lambda_X\bar \psi\tilde \psi\right. \nonumber \\
&&\left.+H^*\left(-g_YY\tan\chi+{g_XQ_X\over
\cos\chi}\right)\lambda_X\tilde
H-\left(-g_YY\tan\chi+{g_XQ_X\over\cos\chi}
\right)\bar\lambda_X\bar{\tilde H}H\right]
\end{eqnarray} 
where $\psi$ and $\tilde \psi$ represent the quarks/leptons and 
the squarks/sleptons, respectively.
Higgs fields $ (H_1, H_2, S)$ are summarized as $H$ and the corresponding
Higgsinos $\tilde H_1$, $\tilde H_2$ and $\tilde S$ are denoted as 
$\tilde H$. 
$Y$ and $Q_X$ stand for $U(1)_Y$ and $U(1)_X$ charges of $\psi$ and $H$.
Noting these modifications, the gaugino-fermion-sfermion vertex can be
expressed by using the following factors,
\begin{eqnarray}
&&Z_i^L(Y, Q_X)=-{1\over \sqrt 2}\left[g_W\tau_3U_{1i}+g_YYU_{2i}+
\left(-g_YY\tan\chi+{g_XQ_X\over \cos\chi}\right)U_{3i}\right], \nonumber \\
&&\overline{Z_i^R}(Y, Q_X)={1\over \sqrt 2}\left[g_YYU_{2i}+
\left(-g_YY\tan\chi+{g_XQ_X\over \cos\chi}\right)U_{3i}\right],
\end{eqnarray}
where we used the left-handed basis for chiral superfields.
It is also useful to define the chargino mass eigenstates for the
calculation of the EDME.
The chargino mass term is given with the matrix form as
\begin{equation}
{\cal L}^{\rm n}_{\rm mass}=-\left(H_1^-, -i\lambda^-_Y\right)
\left(\begin{array}{cc}-\lambda u& \sqrt 2m_Zc_W\cos\beta\\
\sqrt 2m_Zc_W\sin\beta& M_W \\ \end{array}\right)
\left(\begin{array}{c}H_2^+ \\ -i\lambda^+_Y\\ \end{array}\right) +
{\rm h.c.}.
\end{equation}
The mass eigenstates are defined in terms of weak interaction 
eigenstates through the unitary transformations,
\begin{equation}
\left(\begin{array}{c}\tilde \chi_1^+\\ \tilde \chi_2^+\\ \end{array}\right)
=W^{(+)\dagger}\left(\begin{array}{c}H_2^+\\ -i\lambda^+_Y\\
\end{array}\right),  \qquad
\left(\begin{array}{c}\tilde \chi_1^-\\ \tilde \chi_2^-\\ \end{array}\right)
=W^{(-)\dagger}\left(\begin{array}{c}H_1^-\\ -i\lambda^-_Y\\ 
\end{array}\right).
\end{equation}

The effective interaction contributing to the EDME is expressed as
\begin{equation}
{\cal L}_{\rm eff}={i\over 2}{\cal F}_e\bar\psi_e\sigma_{\mu\nu}\gamma_5
\psi_e F^{\mu\nu}, 
\end{equation}
where $\psi_e$ stands for the electron field.
The EDME is given by using this effective coupling ${\cal F}_e$ as
$d_e= {\rm Im}({\cal F}_e)$.
This effective coupling has the contribution from the one-loop
diagrams shown in Fig.1.   
Neutralino contributions come from diagram (a)
and the chargino contribution is represented by (b). 
The effective coupling induced by these diagrams can be calculated as,
\begin{eqnarray}
&&\hspace*{-.8cm}{\cal
F}_e={-e\over16\pi^2}\left[\sum_{\tilde\chi_i^0}
{m_i^n\over M_e^2}
\left\{{g_W m_e \over \sqrt 2 m_W\cos\beta}U_{4i}
\left(Z^L_{2i}(-1,Q_{e_L})~I({m_i^2\over M_e^2})
+\overline{Z^R_i}(2,Q_{e_R})~I({m_i^{n2}\over
M_e^2})\right)\right.\right. \nonumber\\
&&\hspace*{6mm}\left.+\left(Z_{2i}^L(-1,Q_{e_L})\overline{ 
Z^R_i}(2,Q_{e_R}){M_{LR}^{e2}\over M_e^2}+
{g_W^2m_e^2\over 2m_W^2\cos^2\beta}U_{4i}^2
{M_{LR}^{e2\dagger}\over M_e^2}\right)
F({m_i^{n2}\over M_e^2})
\right\} \nonumber\\
&&\hspace*{6mm}\left.+\sum_{\tilde\chi_i^+}{m_i^c\over M^2_\nu}
{m_eg_W^2  \over \sqrt 2 m_W\cos\beta}W_{2i}^{(+)}W_{1i}^{(-)}
J({m_i^{c2}\over M^2_\nu})\right],
\end{eqnarray}
where $m_i^n$ and $m_i^c$ are the $i$-th mass eigenvalue of the neutralino 
and the chargino, respectively.
We derived this formulus in the lepton mass matrix diagonal basis
$m_l=m_l^\alpha\delta_{\alpha\beta}$.
Under this basis the slepton mass matrix is not diagonal 
in general.
For simplicity, these matrix elements may be assumed to follow the
conditions\footnote{
These assumptions are realized in the $N=1$ minimal supergravity with
the supersymmetry breaking in the hidden sector. We do not consider
the lepton flavor violating interaction in this analysis.},
\begin{eqnarray}
&&(M_{LL}^{e2})_{\alpha\beta}=(M_{RR}^{e2})_{\alpha\beta}
=M_e^2\delta_{\alpha\beta},\qquad
(M_{LL}^{\nu 2})_{\alpha\beta}=M_\nu^2\delta_{\alpha\beta}\nonumber \\
&&(M_{LR}^{e2})_{\alpha\beta}=m_l^\alpha(A_l+\lambda^\ast u\tan\beta)
\delta_{\alpha\beta},
\end{eqnarray} 
where $A_l$ is a soft supersymmetry breaking parameter
associated with the charged lepton Yukawa coupling.
Using these assumptions, we adopted the mass insertion approximation 
for the left-right mixing slepton mass in the derivation of eq.(13).
This approximation is expected to be rather good because of the
strong experimental constraints on the flavor changing neutral
current and the smallness of lepton masses $m^\alpha_l$.
In this formulus CP-violating phases are confined in
the left-right mixing slepton mass $M_{LR}^{e2}$ and mixing matrix elements
$U_{ij}$ and $W_{ij}^{(\pm)}$ due to the phases of $A_l$ and $\lambda$
after making gaugino masses real by using the R-transformation.
It should be noted that the superpartner of $\nu_R$ is assumed to be 
heavy enough to decouple from this calculation\footnote{
This may be justified by the fact that the solar neutrino problem
suggests that $\nu_R$s have the large supersymetric mass to make the
seesaw mechanism applicable.} and 
then the chargino contribution becomes very simple.
Kinematical functions $I(r), J(r)$ and $F(r)$ are defined by
\begin{eqnarray}
&&I(r)={1 \over 2(1-r)^2}\left[1+r+{2r \over 1-r}\ln r\right],\nonumber\\
&&J(r)={1 \over 2(1-r)^2}\left[-3+r-{2 \over 1-r}\ln r\right],\nonumber\\
&&F(r)={1\over 2(1-r)^4}\left[1+4r-5r^2+2r(r+2)\ln r\right].
\end{eqnarray}
Our formulus includes the additional parameters besides the ones 
contained in the MSSM formulus :   
$(\tan\beta, M_e^2, M_\nu^2, A_l, M_W, M_Y)$.
They include new gaugino masses $(M_X, M_{YX})$, the kinetic term mixing
parameter $\sin\chi$, the extra $U(1)$ coupling $g_X$ and
charges\footnote{It should be noted that $Q_1+Q_2+Q_S=0$ is satisfied
because of the form of superpotential.}
$(Q_1, Q_2)$ and the $\mu$-term relevant parameters $(\lambda, u)$.
Eq.(13) results in the MSSM formulus\cite{edm,edm2}
by putting these additional parameters zero instead of keeping
$\mu=\lambda u$ constant.

Before proceeding to the numerical analysis it will be useful to make
some remarks on the neuralino dominance condition for the EDME.
In eq.(13) the neutralino mass dependence is extracted as
$r^{1/2}I(r)$ and $r^{1/2}F(r)$  where $r=(m_i^n/M_e)^2$.
The chargino mass contribution is also represented as $r^{1/2}J(r)$ 
where $r=(m_i^c/M_e)^2$.
These functions satisfy the approximate relation 
$r^{1/2}J(r)\sim 4r^{1/2}I(r)\sim 4r^{1/2}F(r)$ at least except for the
region near $r\sim 0$.
Thus the conditon for the neutralino dominace  can be roughly estimated as
\begin{equation}
 \vert A_l +\lambda^\ast u\vert ~{^>_\sim}~
{4 \sqrt 2 M_e^2\over m_W\cos\beta}, 
\end{equation}
where we assumed $M_e\sim M_\nu$.
On the other hand, 
in the present models the vacuum expectation value $u$ of 
the singlet Higgs $S$
is relevant to the extra $Z$ mass in addition to  determining the
$\mu$-scale. The mixing between the ordinary $Z$ and the $U(1)_X$
boson is severely constrained by the precise measurement at
LEP\cite{exp}.
This constraint requires that the mass of the $U(1)_X$ boson is large
enough\footnote{The mixing element of the mass matrix can be small
enough for the special value of $\tan\beta$ and $\sin\chi$.
In such a case this requirement is not necessary to be satisfied and
$u$ may be able to take the rather small value.}
and in such a case the mass eigenvalue of the extra $U(1)_X$ boson is given
as\cite{sy},
\begin{equation}
m_{Z^\prime}^2 \simeq {1\over 2\cos^2\chi}g_X^2(Q_1^2v_1^2+Q_2^2v_2^2+
Q_S^2u^2).
\end{equation}
Although the large $u$ value is necessary to make $m_{Z^\prime}$ large enough,
 the smallness of $\lambda$ may be required 
to give $\mu$ an appropriate value for the large $u$.

Taking account of these aspects, we adopt the parameter set as 
\begin{eqnarray}
&&\tan\beta=1.5  ,\quad \vert A\vert =1500~{\rm GeV},\quad
{\rm Arg}(A)=4\times 10^{-3},\quad \lambda =0.5,\nonumber \\
&&M_e=M_\nu=100~{\rm GeV}, \quad
M_Y=M_X={5\over 3}\tan^2\theta_W M_W, \quad M_{YX}=0,
\end{eqnarray} 
where we assumed the unification relation for the gaugino masses.
For simplicity, $\lambda$ is assumed to be real and then the CP-phase is
included only in $M_{LR}^{e2}$.
As the extra $U(1)_X$ group,  we take the $\eta$-model induced from 
$E_6$ and their charge assingments for 
the relevant fields are listed in Table 1.
\begin{figure}[bt]
\begin{center}
\begin{tabular}{|c||c|c|c|c|c|c|c|c|}\hline
fields & $Q$ & $U^c$ & $D^c$ & $L$ & $E^c$ & $H_1$ & $H_2$ &$S$ \\\hline\hline
$Y$ & ${1\over 3}$  & $-{4\over 3} $ &${2\over 3}$& $-1$ &2 &$-1$ &1 
& 0\\\hline
$Q_{\eta}$& $-{2\over3}$&$-{2\over3}$&${1\over3}$&${1\over3}$&$-{2\over3}$&
${1\over3}$&${4\over3}$&$-{5\over3}$\\\hline
\end{tabular}\vspace{.5cm}\\
{\small {\bf Table 1}
\hspace{.3cm}
The charge assignments of extra $U(1)$s induced from $E_6$.
These charges are normalized as $\displaystyle \sum_{i \in {\bf 27}}Q_i=20$.
Only relevant fields to our study are listed from ${\bf 27}$ of $E_6$.}
\vspace{.5cm}\\
\end{center}
\end{figure}
Here we should remember the allowed region of the $(\mu, M_W)$ plane
obtained from the neutralino and chargino search at LEP\cite{aleph}. 
We confine our study to the $\mu >0$ region so that 
$\mu, M_W~{^>_\sim}~100$~GeV should be satisfied
for $\tan\beta=1.5$. This corresponds to $u^\prime > 2$ for $\lambda=0.5$.
Additionally, if $m_{Z^\prime}~{^>_\sim}~400$~GeV, the rough
estamation with the use of eq.(17) requires $u^\prime~{^>_\sim}~14$. 
Under this parameter setting,
the EDME of this model is plotted in Fig.2 
for $M_W=80$, 180~GeV and $\sin\chi=0$, 0.3 as a function of $u$ where
$u=50(u^\prime+2)$.
As seen from this figure, our parameter set brings the EDME
around the present experimental bound.  
In Fig.3 the ratio of this EDME against the one of the MSSM
is drawn for the comparison with the MSSM.

At first we can see the non-negligible deviation of the EDME from the MSSM
value from Fig.3.
The deviation becomes larger for the larger gaugino mass for $\sin\chi =0$
but the situation is reversed in the case of $\sin\chi =0.3$.
In the large $u$ region where the above mentioned small mixing
constraint is automatically satisfied, the derivation is monotonically 
decreases with $u$.
However, there is $\sim 4$~\% enhancement for $\sin\chi=0$ and also
$\sim 8$~\% enhancement for $\sin\chi=0.3$ around $u^\prime\sim 14$.
This suggests that even in the large $u$ region we may find the
effects of new ingredients $\lambda_X$ and $\tilde S$ in the
neutralino sector on the EDME at the level of $O(10^{-27\sim -28})$ in 
the suitable parameter range.
In the smaller $u$ region this deviation is amplified. In particular,
the gaugino kinetic term mixing shows the larger amplification of the
EDME there.
It is very interesting that the kinetic term mixing effect may be seen 
through the EDME in the present parameter region.
Anyway,  we may fortunately have a chance to find some clue of 
the existence of the extra gauge structure through the study of the EDME
if the experimental bound is improved by order one.

Some brief comments are useful to be ordered on the parameter
dependence of the 
EDME on $\lambda$ and $\tan\beta$ here.
Although there is no significant difference in the absolute value of
the EDME between $\tan\beta=1.5$ and $\tan\beta=50$, 
there can be seen its slight dependence on $\lambda$.
For example, for the case of $\lambda=0.07$, we can find some
enhancement of the EDME for $M_W=80$~GeV compared with $\lambda=0.5$.
However, we cannot see such a substantial change for $M_W=180$~GeV.
On the ratio $R$ of the EDME, we cannot see the significant
dependence of the EDME on both of $\lambda$ and $\tan\beta$ at least in the
large $u$ region (for example, $u^\prime ~{^>_\sim}~ 14$).   
In the present analysis we assumed $\lambda$ is real and only the
origin of the CP violating phase is $A_l$ in $M_{LR}^{e2}$.
Since $\lambda u\tan\beta$ in $M_{LR}^{e2}$ is irrelevant to the EDME
under such an assumption,
the $\lambda$ and $\tan\beta$ dependence come only through the 
diagonalization matrices $U$ and $W^{(\pm)}$ 
of the neutralino and chargino mass matrices.
Because of this reason, the ratio of the EDME of the $\eta$-model 
against the MSSM is considered not to be sensitive to these 
parameters except for the small $u$ region.
If we introduce the phase into $\lambda$ which seems to be more
general situation, the behavior of the EDME is expected to be more
complicated. 
This aspect will be beyond the scope of the present study and we will
present it elsewhere.

In summary we investigated the EDME in the 
extra $U(1)$ models which can potentially solve the $\mu$-problem. 
We pointed out that these models may be distinguished
from the MSSM through the measurement of the EDME and we showed it
concretely for a certain
parameter region in the $\eta$-model derived from $E_6$. 
It is noticable that the abelian gaugino kinetic term 
mixing can have rather large effects on the value of EDME.
It may be possible to find some clue of the extra gauge structure 
by investigating the EDME if the experimental bound is improved.
Our numerical study in this paper has been done in the rather
restricted parameter region of a typical model 
where the neutralino dominance is expected.
However, it will be necessary to investigate this process in more wide
parameter region of various extra $U(1)$ models 
to give the basis of the experimental study.
It will be also useful to do the combined study with other 
process like $\mu\rightarrow e\gamma$\cite{s} where the neutralino 
sector is expected to play the important role.
We will present them elsewhere.
\vspace{.5cm}

This work is partially supported by a Grant-in-Aid for Scientific
Research from the Ministry of Education, Science and Culture
(\#08640362).


\newpage
\noindent
{\Large\bf Figure Captions}
\vspace{1cm}\\
{\Large\bf Fig. 1}
\vspace{.3cm}\\
One-loop diagrams for the EDME.
Fig.(a) represents the neutralino contribution and Fig.(b) 
represents the chargino contribution. CP violating phases are included 
in the slepton mass insertion which is expressed by $\bullet$ and
vertex factors ${\cal V}_i$.  It should be noted that the chirality
flip occurs at the verteces and/or the internal line. 
\vspace{1cm}\\
{\Large\bf Fig. 2}
\vspace{.3cm}\\ 
The EDME as a function of $u$ in the $\eta$-model.
The vertical axis stands for $d_e\times 10^{26}~{\rm e~ cm}$ and 
the horizontal axis $u^\prime$ should be understood as
$u=50(u^\prime +2)$.
Each line corresponds to the parameter settings for $(M_W,~ \sin\chi)$ 
and their values are taken as $A(80,~ 0)$, $B(180,~ 0)$, 
$C(80,~ 0.3)$ and $D(180,~ 0.3)$.
\vspace{1cm}\\
{\Large\bf Fig. 3}
\vspace{.3cm}\\
The ratio $R$ of the EDME of the $\eta$-model against the MSSM
as a function of $u$. $R=d_e^\eta(M_W,~\sin\chi)/d_e^{\rm MSSM}(M_W)$.
Each line corresponds to the same parameter settings in Fig.2.

\newpage
\pagestyle{empty}
\setlength{\unitlength}{1mm}
\begin{center}
\begin{picture}(72,50)(0,0)
\thicklines
\put(35,3){\bf (a)}
\put(63,44){${\cal A}_\mu$}
\put(12,44){$e_L$}
\put(63,10){$e_R^c$}
\put(25,43){${\cal V}_{e_L}$}
\put(50,11){${\cal V}_{e_R^c}$}
\put(35,43){$\tilde e_L$}
\put(56,26){$\tilde e_R^\ast$}
\put(28,31){$\tilde\chi^0$}
\put(40,19){$\tilde\chi^0$}
\put(8,40){\vector(1,0){10}}
\put(18,40){\line(1,0){10}}
\multiput(28,40)(5,0){5}{\line(1,0){3}}
\put(28,40){\line(1,-1){6}}
\put(40,28){\vector(-1,1){6}}
\put(40,28){\vector(1,-1){6}}
\put(38,27){$\bigoplus$}
\put(52,16){\line(-1,1){6}}
\multiput(52,16)(0,5){5}{\line(0,1){3}}
\multiput(53,40)(4,0){5}{\oval(2,3)[t]}
\multiput(55,40)(4,0){5}{\oval(2,3)[b]}
\put(52,16){\line(1,0){10}}
\put(72,16){\vector(-1,0){10}}
\put(45,40){\circle*{3}}
\end{picture}\hspace{1cm}
\begin{picture}(72,50)(0,0)
\thicklines
\put(35,3){\bf (b)}
\put(63,10){${\cal A}_\mu$}
\put(12,44){$e_L$}
\put(63,44){$e_R^c$}
\put(33,44){$\tilde \nu_{e_L}$}
\put(44,44){$\tilde \nu_{e_L}^\ast$}
\put(41,40){\circle*{3}}
\put(56,26){$\tilde\chi^-$}
\put(31,26){$\tilde\chi^+$}
\put(25,35){${\cal V}_{e_L}$}
\put(54,36){${\cal V}_{e_R^c}$}
\put(8,40){\vector(1,0){10}}
\put(18,40){\line(1,0){10}}
\multiput(28,40)(5,0){5}{\line(1,0){3}}
\put(52,16){\vector(-1,1){16}}
\put(28,40){\line(1,-1){8}}
\put(42,23){$\bigoplus$}
\put(52,16){\vector(0,1){15}}
\put(52,28){\line(0,1){12}}
\put(52,40){\line(1,0){10}}
\put(72,40){\vector(-1,0){10}}
\multiput(53,16)(4,0){5}{\oval(2,3)[t]}
\multiput(55,16)(4,0){5}{\oval(2,3)[b]}
\end{picture}

\vspace*{3cm}
{\Large\bf Fig.1}
\end{center}
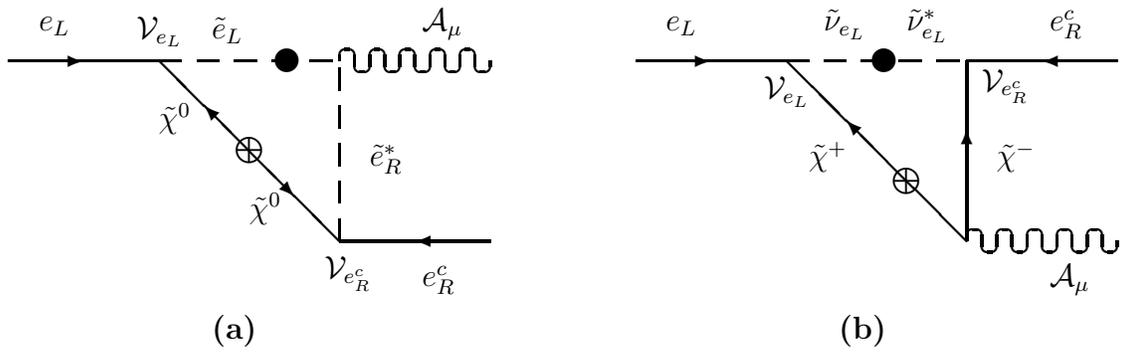
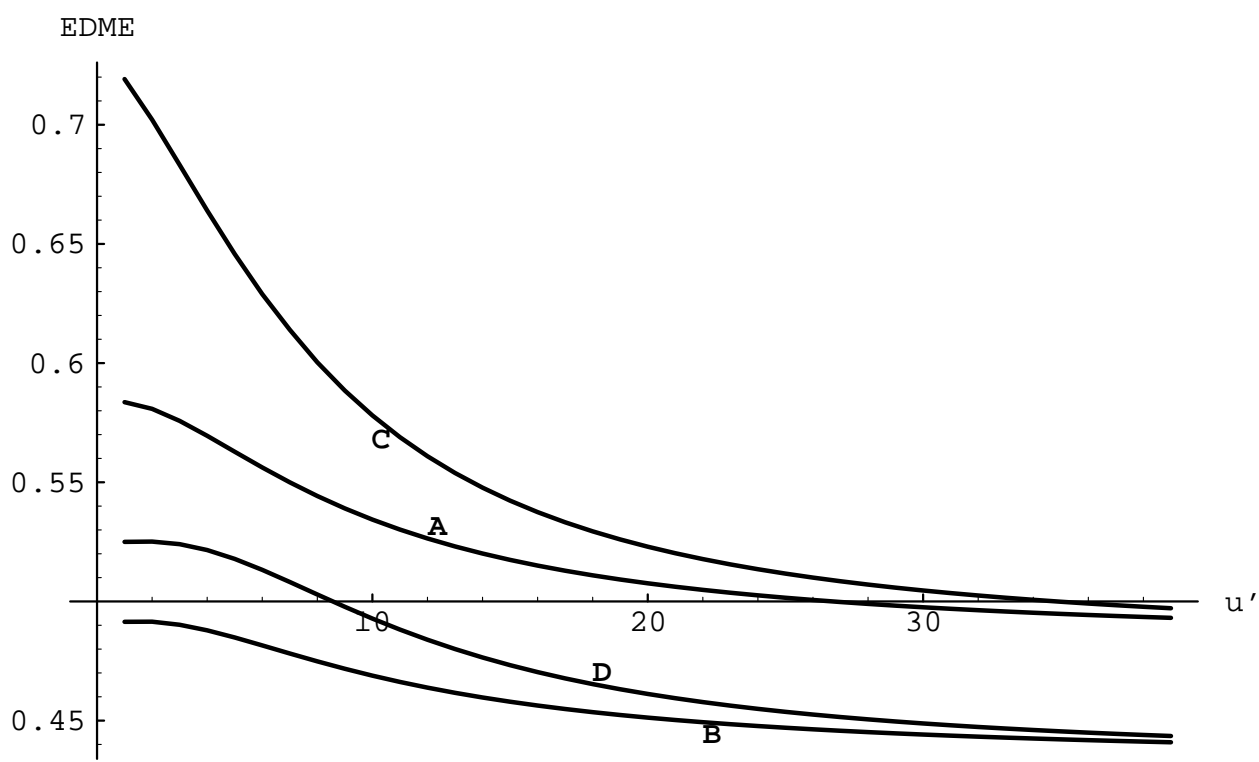
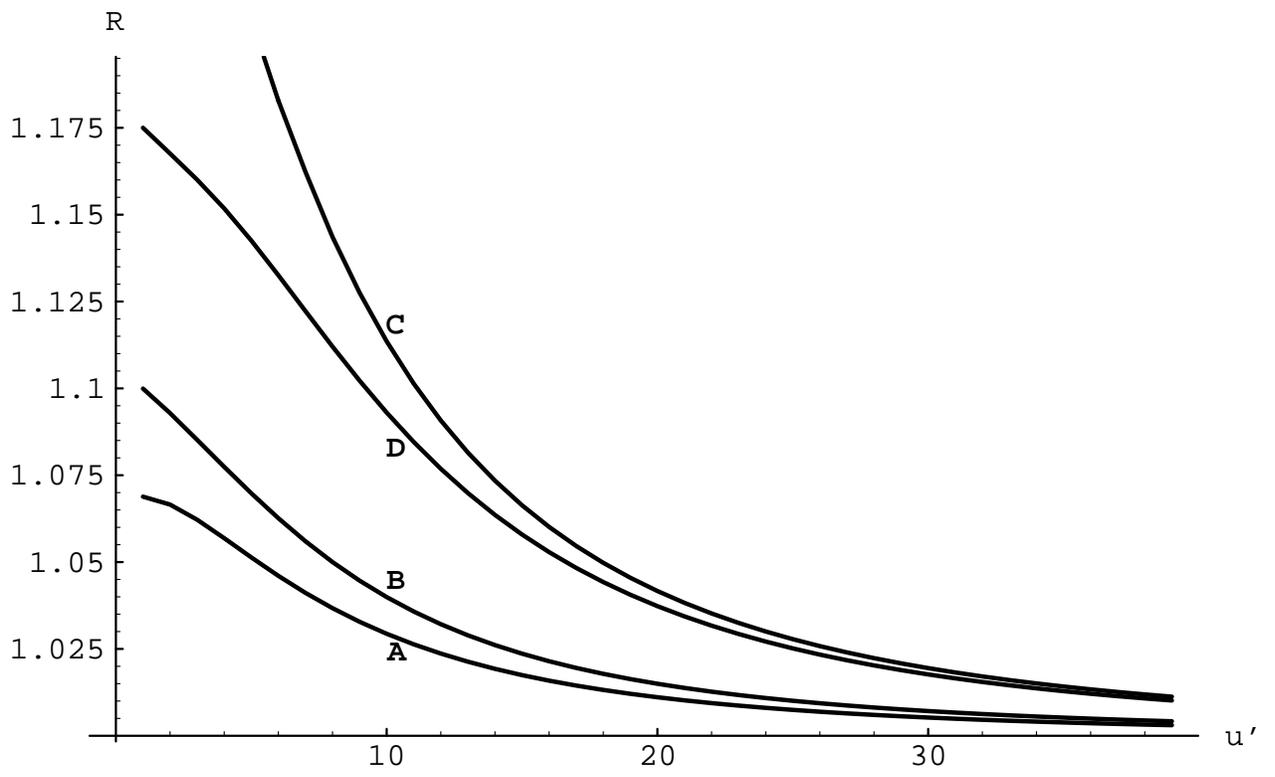
\end{document}